# Shells in CO$_2$ clusters

John W. Niman, Benjamin S. Kamerin, Vitaly V. Kresin

Department of Physics and Astronomy, University of Southern California, Los Angeles, California 90089-0484, USA

Jan Krohn, Ruth Signorell,

Laboratory of Physical Chemistry, ETH Zürich, Vladimir-Prelog Weg 2, CH-8093 Zürich, Switzerland

Roope Halonen,

Center for Joint Quantum Studies and Department of Physics, School of Science, Tianjin University, 92 Weijin Road, Tianjin 300072, China

Klavs Hansen,[1]

Center for Joint Quantum Studies and Department of Physics, School of Science, Tianjin University, 92 Weijin Road, Tianjin 300072, China

and

Lanzhou Center for Theoretical Physics, Key Laboratory of Theoretical Physics of Gansu Province, School of Physical Science and Technology, Lanzhou University, Lanzhou 730000, China.

## 1 Introduction

One of the most striking phenomena associated with clusters is the strong non-monotonic variation of their properties with size. Such finite size effects have been observed in a number of different types of clusters, composed by materials as diverse as atoms of noble gases[1] or simple metals[2, 3], as well as in the all-carbon fullerenes[4]. The variations reflect the shell structure of the clusters, which can be of electronic nature[2, 5] or arising out of the packing of atoms[1, 6].

---

[1]E-mail: klavshansen@tju.edu.cn. Homepage: http://www.klavshansen.cn/

The shell structures in these systems were discovered in molecular beams, manifested in the highly irregular variation of the abundances with cluster size. Shell structure appears in mass abundance spectra because the size-to-size intensity variations reflect the cluster binding energies. The connection between cluster stabilities (*i.e.*, binding energies) and their abundances is, however, not elementary, and cannot be understood as simple equilibrium distributions with the temperature set by the source temperature. In many cases involving cluster beams one finds that the underlying process that shapes the size-to-size abundance variations is that of evaporative cascades: internally excited (hot) clusters undergo a series of evaporation steps resulting in a detected population where the high cluster intensities reflect lower than average evaporation rates and vice versa. The high intensity clusters, often labeled "magic numbers," are frequently assigned a special stability. This is, however, a simplified view that will only hold in special situations, as the general theory below shows, and this must be taken into account in the quantitative analysis of cluster binding energies extracted from such spectra.

Importantly, just a few evaporative steps are sufficient for the population patterns to acquire the shapes that characterize the species [7, 8]. These shapes make it possible to use measurements of relative abundances to extract quantitative information about the monomer-by-monomer variations of cluster binding energies with size. The connection between abundances and binding energies was derived in Ref. 8 and is discussed at length in Ref. 9. It has been applied previously to analyze mass spectra of sodium clusters [10], for which a dedicated experiment unambiguously confirmed the shell energy amplitudes derived from the abundance spectra [11]. It has also been used for clusters of both light and heavy water [12, 13], quantifying in particular the excess stability of the $N = 21$ protonated cluster that gives rise to the well-known abundance peak at that size. Finally, it was applied to find the energy amplitudes of the packing shells which shape the rare gas cluster mass spectra [14].

This work applies the analysis to a large number of abundance spectra of $CO_2$ clusters. The experiments were performed to study nucleation in supersaturated gases [15], but are equally useful for the analysis here, in particular because the wide range of nucleation and detection conditions employed in the measurements offers an uncommonly rich data set. We will demonstrate that the analysis of the mass spectra reveals that highly universal patterns are present in all observed distributions. The derived stabilities are assigned to the neutral clusters produced in the beam.

The plan of the remainder of the paper is as follows. First a brief description of the experimental procedure is given. Then the theory of the formation of the abundance spectra and the analysis are described. Next the dissociation energies extracted from the analysis are given, followed by a section where these values are discussed in terms of packing shell structure. The results are discussed and summarized in the concluding section.

## 2   Experimental procedure and results

The experimental equipment has been described in detail in Ref. 16 where it was employed for nucleation studies[15–17], and only a brief summary is given here.

Figure 1 shows a schematic drawing of the setup. Clusters were produced by co-expansion of $CO_2$ with argon, which acted as a carrier gas, through a pulsed Laval nozzle with a throat diameter of 4.1 mm. The gas expanded from a stagnation pressure of $p_0 \approx 8 \times 10^4$ Pa and room temperature. The $CO_2$ mole fraction before expansion was varied between 0.38 % and 5.02 %.

At a distance $\ell$ after the nozzle, the core of the expansion was sampled with a skimmer and the clusters were single-photon ionized by 13.8 eV (89.8 nm) photons generated with a home-built tabletop vacuum ultraviolet (VUV) laser. The laser operates with 2-color-4-wave mixing in an expanding krypton gas at a repetition rate of 20 Hz. By varying the distance between the nozzle exit and the skimmer, the beam could be sampled at different times in the post-nozzle flow. After ionization the clusters were were accelerated to 30 keV and the mass spectra were measured in a linear time-of-flight mass spectrometer (TOFMS) equipped with a microchannel plate (MCP) detector. The resolution of the TOFMS was 800 at $m/z = 12000$ u. No sign of multiply charged clusters were observed in the relevant size range. The appearance size for doubly charged clusters is $N = 44$ [18] and if they had been present odd-numbered cluster sizes would have been easily seen as nominally half-integer mass peaks.

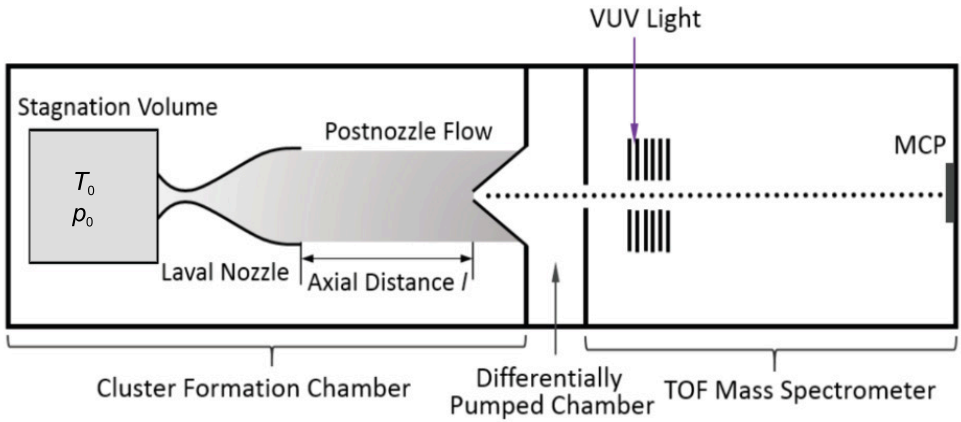

Figure 1: Outline of the experimental setup. Clusters were generated by the expansion of a mixture of $CO_2$ and argon as a cooling gas, photoionized, and detected by a time-of-flight mass spectrometer, as described in the text. Figure adapted from Fig. 1 of Ref. 19 with permission from the PCCP Owner Societies.

Figure 2 shows three examples of mass spectra recorded with different $CO_2$ mole fractions and nozzle-TOFMS separations. These mass spectra were obtained from the raw time-of-flight data by applying background subtraction and rescaling, as described in the Electronic Supplementary Information (ESI). The variation of the average cluster size with source parameters is discussed in Ref. 15, and since average sizes are not relevant for the analysis here, we will refrain from a detailed description of this aspect.

The $(CO_2)_N^+$ distributions show a clearly visible structure with periodic intensity modulations with a period on the order of 10 monomers. The pattern seen in the figure is reproducible for clusters larger than approximately 130 molecules. It has been observed previously [15, 20] and ascribed to shell closings in cuboctahedral cluster structures. Sim-ilar variations have been seen in anionic clusters [21]. In the present paper the focus is on the important information about cluster structure and in particular about the magnitude of the underlying stability variations that can be extracted from these persistent patterns.

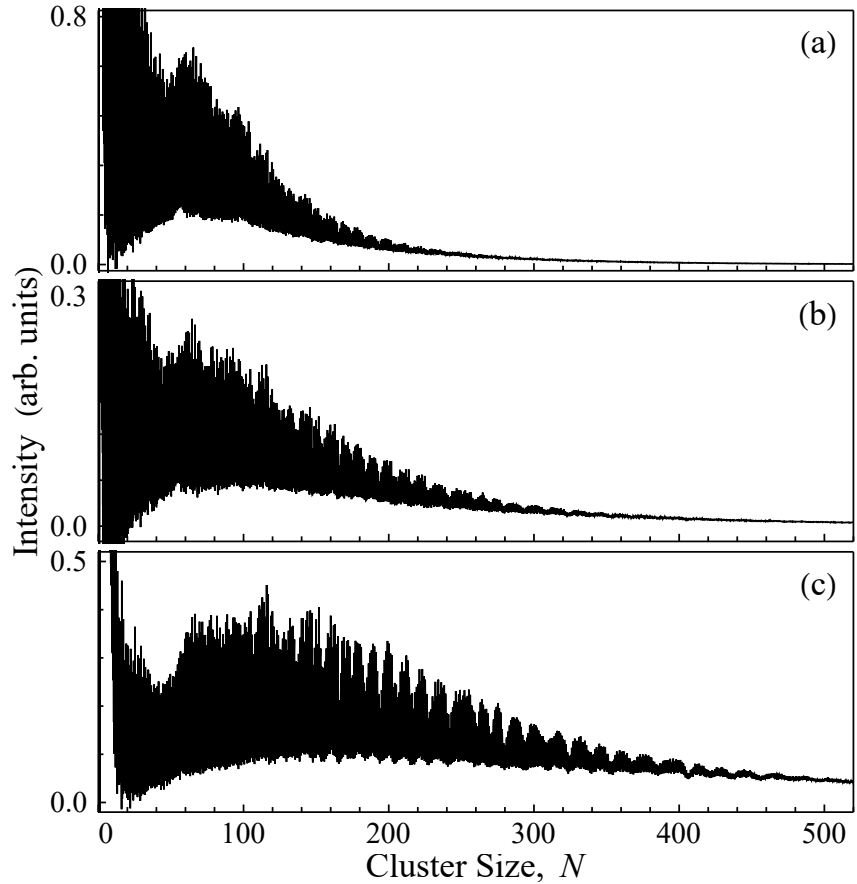

Figure 2: Three representative cluster mass spectra with different $CO_2$-argon mixtures and nozzle-TOFMS separations: (a) 1.54 % $CO_2$ mole fraction and 323 mm nozzle-ionization distance, (b) 1.54 % $CO_2$ and 403 mm distance, (c) 3.85 % $CO_2$ and 403 mm distance. Spectra are reproduced from data reported in Ref. 15.

## 3  Data analysis

The minima in the mass spectra, $N_{\min}$, are well defined, and for a first approximate picture of the stability pattern the cube roots of their positions are plotted vs. their number of appearance. Such plots are shown in Fig. 3 for the three spectra shown in Fig. 2 (integrated as described below). The nearly equidistant spacing, here with approximately ten dips for each unity increment of $N_{\min}^{1/3}$, is a signature of shell structure [6]. The numerical value of the spacing indicates that the structure is face centered cubic, either cuboctahedral (truncated *fcc*), as already suggested in Ref. 20, or distorted (octahedral) *fcc*[22].

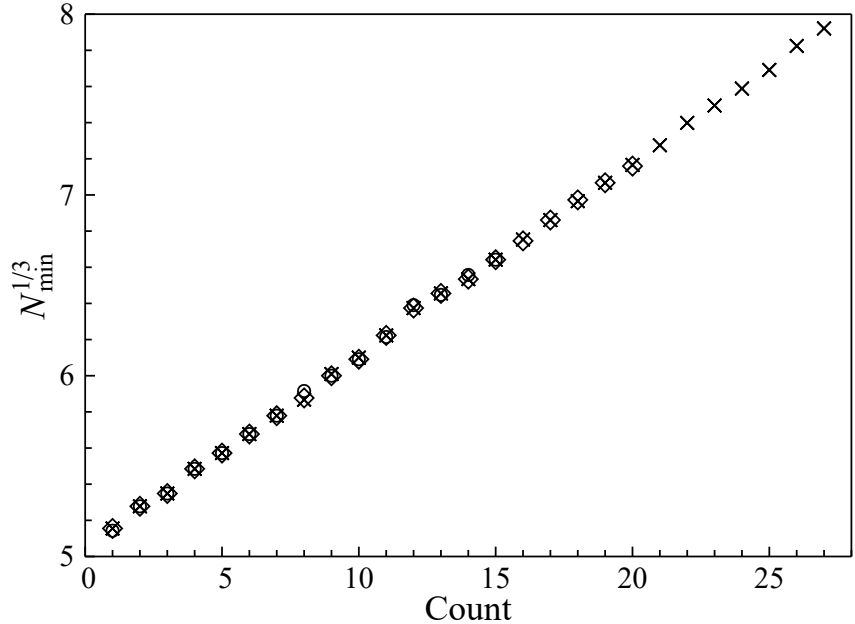

Figure 3: The cube root of cluster sizes $N_{\min}$ corresponding to abundance minima, plotted in the order of their appearance in the abundance spectra of Fig. 4 below. Circles, squares, and crosses correspond to spectra labeled (a), (b), and (c), respectively. The count included in this plot starts at $N_{\min} = 130$.

After confirming the assignment of the intensity variations to shell structure, two questions arise. First, one may inquire about the precise *location* of the shells (or subshells), because these are almost certainly not coincident with the abundance spectra minima. The second question concerns the *energy amplitude* of the shell modulation that is manifested in the abundance spectra. Both of these questions will be answered by application of the theory mentioned in the Introduction and given in detail in Refs. 8 and 9.

The analysis of the mass spectra begins with an integration of the individual mass peaks. As described in detail in the ESI, this involves identifying and subtracting a constant baseline and incorporating a smooth correction for the mass scaling and photoionization efficiency. Following these steps, the midpoints between mass peaks are identified and the intensity between these is integrated. This yields the ion intensities $I_N$ as a function of cluster size $N$.

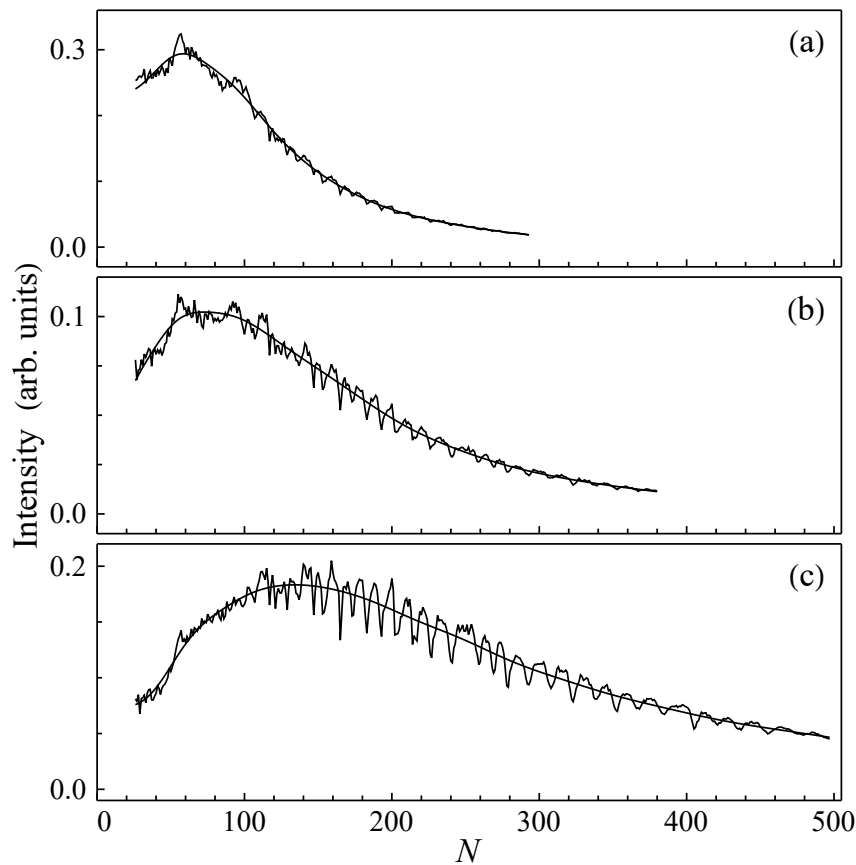

Figure 4: Integrated abundance spectra $I_N$ derived from the mass spectra in Fig. 2, and their smooth envelope functions $\tilde{I}_N$.

Spectra recorded under different source conditions are made up of a smooth envelope function modulated by the abundance variations. The latter are shaped by the evaporative losses and carry the information that is of interest here. The smooth envelope function, in contrast, is shaped by the precise parameters of the cluster source. In order to extract the evaporative abundance variations from the spectra recorded under different source conditions, the envelope function is determined for each spectrum and divided out.

These envelope functions, denoted $\tilde{I}_N$, were calculated by iterative convolution of the integrated mass spectra with Gaussian functions, as described in the ESI. Examples of the resulting envelope functions are shown in Fig. 4, plotted together with the individual peak intensities.

After division of the intensity spectra by $\tilde{I}_N$, the thus normalized abundance variation ratios $I_N/\tilde{I}_N$, referred to as stability functions, oscillate around unity. The outcome of this analysis for the three sample spectra from Figs. 2 and 4 with their own envelope functions is shown in Fig. 5, together with the mean stability function of all experimental spectra.

Strikingly, the stability functions derived from all the mass spectra are practically identical in their overlapping regions for values above $N \approx 130$. The good agreement between stability functions extracted from mass spectra produced under a range of different conditions allows us to conclude that they reflect inherent cluster properties, consistent with the hypothesis that they are shaped by evaporative events after production. In contrast, the envelope functions differ widely for different source conditions, as expected from the correspondingly different nucleation parameters [15, 23]. Indeed, although it cannot be excluded that clusters may undergo some additional collisions even in the post-skimmer collimated flow, strong size-to-size intensity oscillations are a hallmark outcome of evaporative processes.

The next step in the analysis is to relate the stability function, $I_N/\tilde{I}_N$, to the cluster energies. The function is shaped by the clusters' evaporative activation energies, $D_N$, which are the main determining factors for the speed of evaporation that can have a non-monotonic size-to-size variation. They can be taken to be identical to the cluster dissociation energies. This identification holds for a molecule-cluster potential without any barrier to attachment, which can be safely assumed for $CO_2$.

As mentioned above, the number of molecules which is required to have evaporated in order to apply the analysis is small. For the present systems a few evaporative steps suffice[8]. This means, in particular, that conclusions about nucleation[15] drawn from the data are not in any conflict with the present analysis and vice versa.

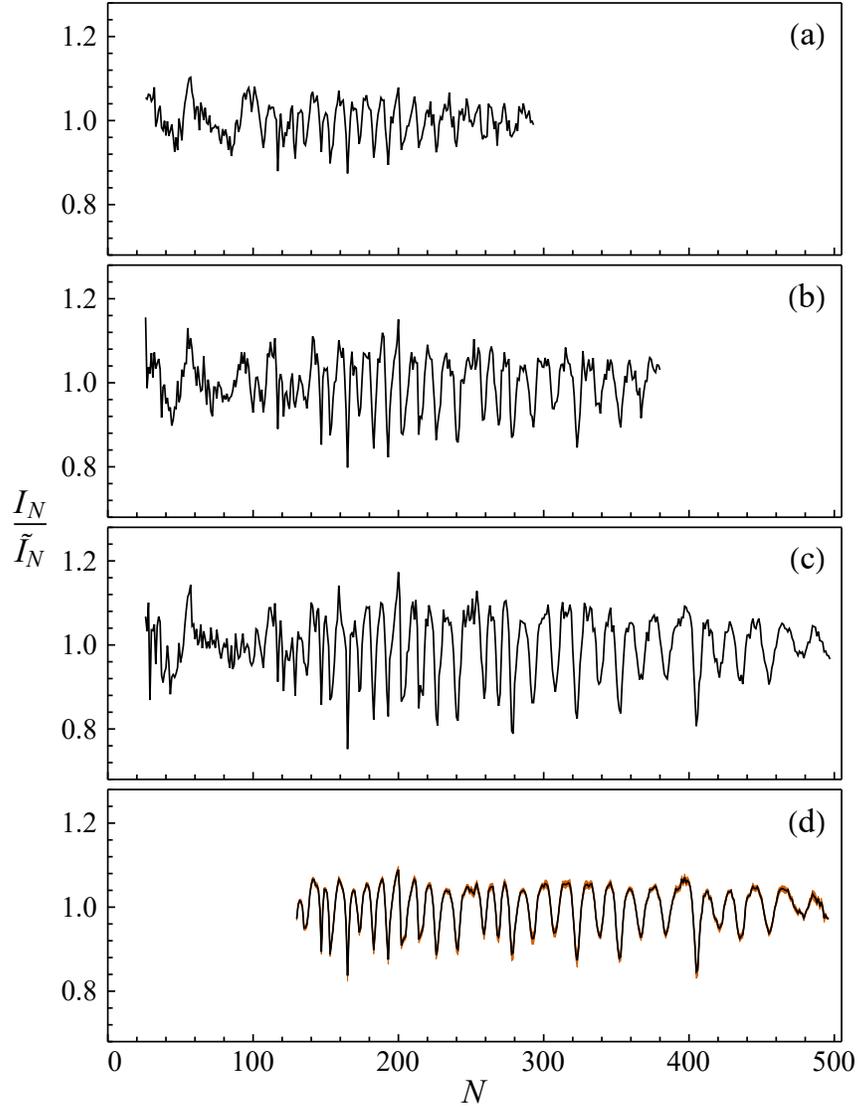

Figure 5: Top three panels (a–c): stability functions of the three spectra shown in Fig. 4. Bottom panel (d): the average stability function derived from all the mass spectra in the experimental data set (black line) with the standard error of the mean indicated by a brown field.

The analysis in Ref. 8 established the following relation between the stability functions and cluster dissociation energies:

$$\frac{I_N}{\tilde{I}_N} = \frac{D_N + D_{N+1}}{2\tilde{D}_N} + \frac{C_N}{\ln(\omega_N t)} \frac{D_N - D_{N+1}}{\tilde{D}_N}. \tag{1}$$

Here $\tilde{D}_N$ is the part of the dissociation energy which varies smoothly with cluster size.

It is analogous to the energy of Thomson's drop model[9] and to the liquid drop part of nuclear[24, 25] and metal cluster[26–28] binding energies. It should be emphasized that in spite of the name, the applicability of such parametrization is not restricted to liquid phase particles: the essential point is that the energy has a smoothly varying size dependence. $C_N$ is the vibrational heat capacity of the cluster (in units of $k_B$) for which the bulk heat capacity of solid $CO_2$, scaled to the cluster size $N$, is used. Additional small corrections for the microcanonical nature of the process[29] and the overall translational and rotational degrees of freedom are included (see the ESI for details).

The quantity $G_N = \ln(\omega_N t)$ is referred to as the Gspann parameter[30, 31]. Here $t$ is the time elapsed between the production of the clusters and the completion of the mass selection in the acceleration stage of the TOFMS. The factor $\omega_N$ is the frequency prefactor in the expression for the unimolecular rate constant that describes the statistical process of monomer loss from the clusters. Its value can be estimated from molecular properties, but a simpler procedure is to extract it from the bulk vapor pressure together with the molecular area from the measured bulk density. The procedure is described in detail in the ESI. For the cluster sizes studied here, $G_N$ is found to vary between 32 and 35.

With these two parameters known, the difference equation (1) can be solved numerically. We rewrite it, ignoring the small difference between $\tilde{D}_N$ and $\tilde{D}_{N+1}$, as

$$\frac{D_N}{\tilde{D}_N} = \frac{1}{\frac{C_N}{G_N} + \frac{1}{2}} \left[ \frac{I_N}{\tilde{I}_N} + \frac{D_{N+1}}{\tilde{D}_{N+1}} \left( \frac{C_N}{G_N} - \frac{1}{2} \right) \right], \qquad (2)$$

and solve this iteratively. The value of $D_N/\tilde{D}_N$ for the largest size in a spectrum is required as input. Regardless of the precise value of this starting value, the procedure converges to a stable set of dissociation energies for lower $N$. However, the speed of convergence depends on the chosen starting value. To optimize the convergence speed we varied this value by minimizing the deviation from unity of the resulting set of solutions for all sizes $N$, as described in the ESI. In all cases these optimized values were consistent with values extracted from the procedure applied to other spectra in overlapping mass regions, confirming the soundness of the procedure.

# 4 Dissociation energies

The ratios $D_N/\tilde{D}_N$ derived from the spectra in Fig. 5 are displayed in Fig. 6(a–c), and Fig. 6(d) shows the average of all spectra. The variations of the $D_N/\tilde{D}$ values follow those of the stability functions with some important differences.

First of all, the amplitudes of the dissociation energy variations are much smaller than those of the stability functions, due to the large heat capacity factor multiplying the energy differences. This will amplify measured abundance variations very strongly, and more so the larger the clusters. The effect is known and has been observed previously (see Ref. 32 for an extreme case of this amplification). Conversely, this means that when clusters of different sizes are observed to display abundance variations of a similar magnitude, the underlying energy variations are actually larger for the smaller clusters. This is a direct consequence of the above equations but is worth highlighting.

The second important difference is that the maxima and minima of the structure function curves and the energy curves are shifted relative to each other. This is likewise a consequence of the fact that the second term in Eq. (1) is much larger than the first, and that high abundances therefore occur where the dissociation energy *experiences a drop with increasing size* and *not* where it is high.

It is useful to convert the results to absolute energies. This is done by multiplication with the Thomson liquid drop energies, determined by bulk parameters as

$$\tilde{D}_N = A - \frac{2}{3}BN^{-1/3}, \qquad (3)$$

where $A$ is the bulk binding energy per molecule, and $B$ is related to the surface tension, $\gamma$, via

$$BN^{2/3} = 4\pi r_0^2 N^{2/3}\gamma, \qquad (4)$$

where $r_0$ is the molecular averaged radius, defined by the density.

The experimental enthalpy of sublimation[33], $27.2 \pm 0.4$ kJ/mol (0.28 eV), is used for the value of $A$. This is not precisely the same quantity as $A$, but the difference involves only a small difference of thermal energies which can be ignored for the present purpose. The value is close to the one found theoretically in Ref. 23 where macroscopic parameters were used to adjust the interaction potentials and simulations were performed for finite excitation energy clusters in similar size ranges.

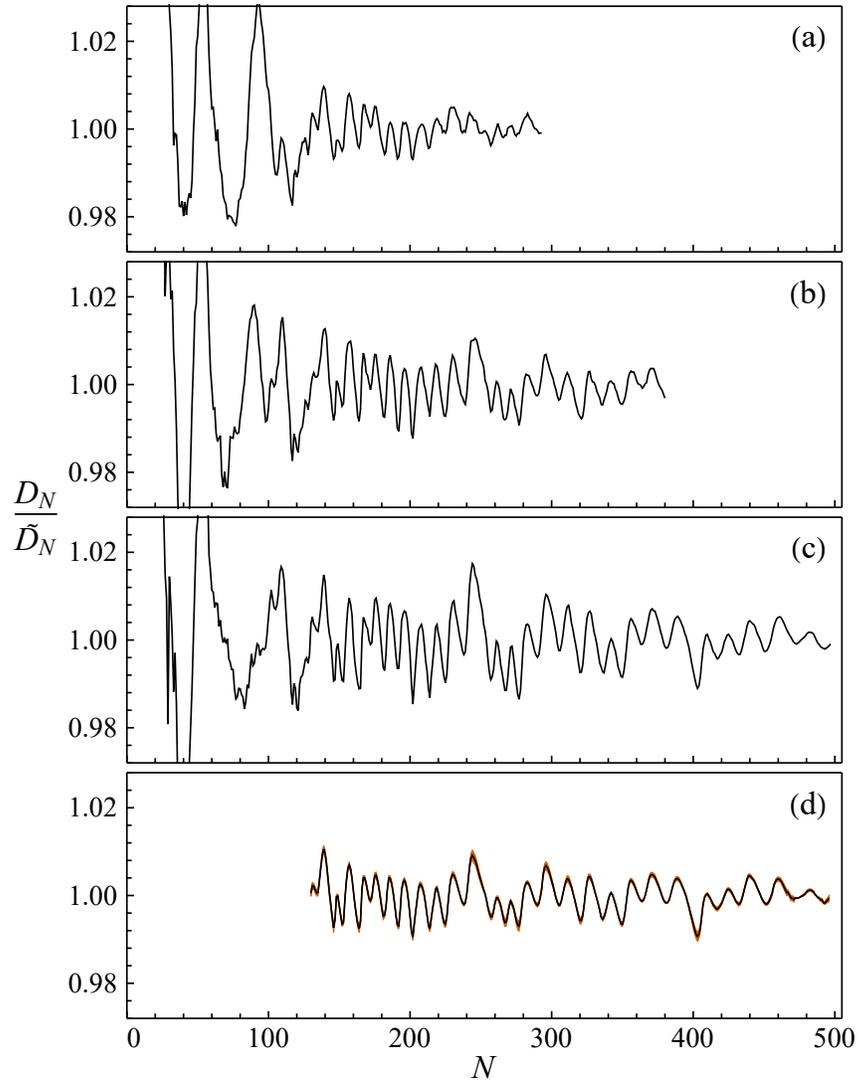

Figure 6: Top three panels (a–c): dissociation energy ratios calculated from the stability functions in Fig. 5(a–c). Bottom panel (d): dissociation energies averaged over the full data set (black line) with the standard error of the mean given by the brown field. Note the large difference between the scales of the variation of the stability functions and the dissociation energies. This is due to the large value of the ratio $C_N/G_N$ for these cluster sizes.

No reliable data have been found for the surface energy of solid $CO_2$, and we will use the relation

$$B = \frac{2}{3}A = 0.188 \text{ eV}, \tag{5}$$

which has been found to give fair estimates for a number of substances, including van der Waals bound solids [9]. The 0 K value derived from Ref. 23 is closer to $B \approx A$. The difference in the cluster dissociation energies between these two variants amounts to a shift downward of about 0.01 eV with very little effect on the relative variations, and can be ignored without any major loss of precision. The dissociation energies calculated from the results in Fig. 6(d) by using these parameters are shown in Fig. 7.

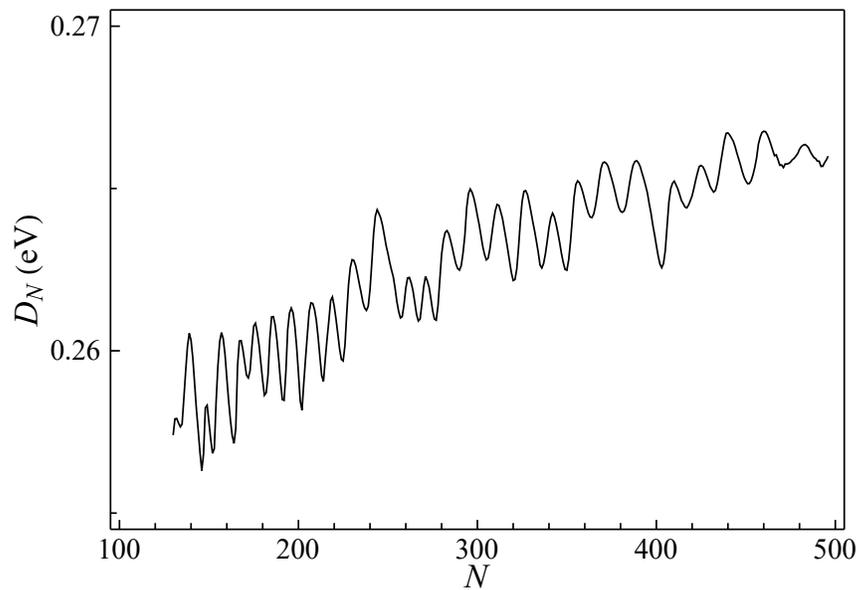

Figure 7: Cluster dissociation energies calculated from the ratios shown in Fig. 6(d) by using Eq. (3) for $\tilde{D}_N$.

## 5 Shell structure

As emphasized above, the positions of shell closings do not coincide with the abundance ($I_N$) maxima. They also are not necessarily given by the maxima in the dissociation energies, $D_N$, at least at finite temperatures where shell closings tend to spread out over more sizes with increasing amount of thermal excitations. From experimental results on the shell structure of sodium clusters[11] it was concluded that at finite temperatures the shell structure's prototypical sawtooth variation of dissociation energies with cluster size becomes rounded, and the location of shell closings in the presence of such rounding can be identified with the point of steepest descent in the curve of $D_N/\tilde{D}_N$ vs. $N$. A discussion of this question, applied to experiments on clusters of rare gas atoms, can also be found in Ref. 34. We will likewise identify the steepest slope with the shell closing.

The sequence of shell and subshell closings of $(CO_2)_{N \gtrsim 130}$ clusters is determined according to this prescription from the data in Fig. 6(d). Details of the numerical procedure are described in the ESI. The order of occurrence of subshell closings is quantified with the subshell closing index $k' = Fk$, where $F$ is the number of facets on a closed shell cluster. This subshell index $k'$ accounts for individual facets between closed shells given by the index $k$. The subshell indices for the obtained closings are tentatively assigned by associating the observed closing at $N_s = 143$ with the ideal cuboctahedral cluster of $k = 4$ (i.e., $N = 147$) with the index $k' = 14 \times 4 = 56$ as illustrated in Fig. 8(a). As shown in Fig. 9, the cube roots of $N_s$ lie on a straight line when plotted against the assigned $k'$. It is worth pointing out that this identification of the shells structure is more precise and rigorous than the approximate one based on abundance minima illustrated in Fig. 3. The obtained slope of the $N_s^{1/3}$ vs. $k'$ curve, $0.1060 \pm 0.0002$, is in good agreement with the coefficient for cuboctahedral filling of $k'$ subshells (see the ESI and Refs. 20 and 35 for a detailed description of these subshell closings):

$$N_s^{1/3} \approx 0.1067(k' - 7) \tag{6}$$

Using Eq. (6) as the regression equation also minimizes the y-axis intercept ($N_s^{1/3} = 0.1060(k' - 7) + 0.028$). This result further supports the aforementioned assignment of the subshell closings according to a cuboctahedral cluster.

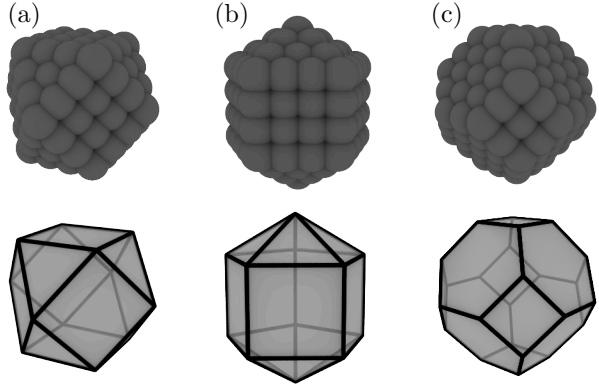

Figure 8: Shell closings for (a) cuboctahedron with 14 facets ($k = 4$, $N = 147$), (b) regular Ino decahedron with 15 facets ($k = 4$, $N = 147$), and (c) truncated octahedron with 14 facets ($k = 3$, $N = 201$). Detailed discussion and analysis about the geometries of these structures and their (sub)shell closings is given in the ESI.

The very good agreement of the positions of the subshell closings with those predicted by cuboctahedral structures notwithstanding, it is still of interest to compare the data with alternative structures. In the previous structural analysis by Negishi et al. [20], fcc cubic, octahedral and icosahedral geometries were included in addition to cuboctahedron. From this set of different structures it is clear that only a cuboctahedral geometry is able to capture the observed $N_s^{1/3}$ vs. $k'$ behavior of $CO_2$ clusters. However, their analysis ignored structures such as Ino (or Marks) decahedra and truncated octahedra illustrated in Figs. 8(b) and 8(c), respectively. These are generally plausible alternatives for larger clusters[36–39]. For this reason we extended the geometric analysis by Negishi et al. [20] (and Näher et al. [40]) to truncated octahedral and Ino decahedral clusters. The slopes of the $N_s^{1/3}$ vs. $k'$ curves are expected to be 0.100 for the Ino decahedra and 0.110 for the truncated octahedra, which are substantially closer to the cuboctahedral value than any other structure considered by Negishi et al. [20]. To show the general applicability of the used structure identification via cluster energy variations, we have carried out test calculations based on a peeling-off process of the least bound monomers. In short, these calculations fully support the adequacy of the geometric analysis of cluster packing used here and in previous studies. The procedure and the obtained results are discussed in detail in the ESI. The conclusion is that the two alternative structures both give a clearly worse fit of slope of the experimental data, and that the presented sequence of shell closings of $CO_2$ is best characterized by the cuboctahedral geometry.

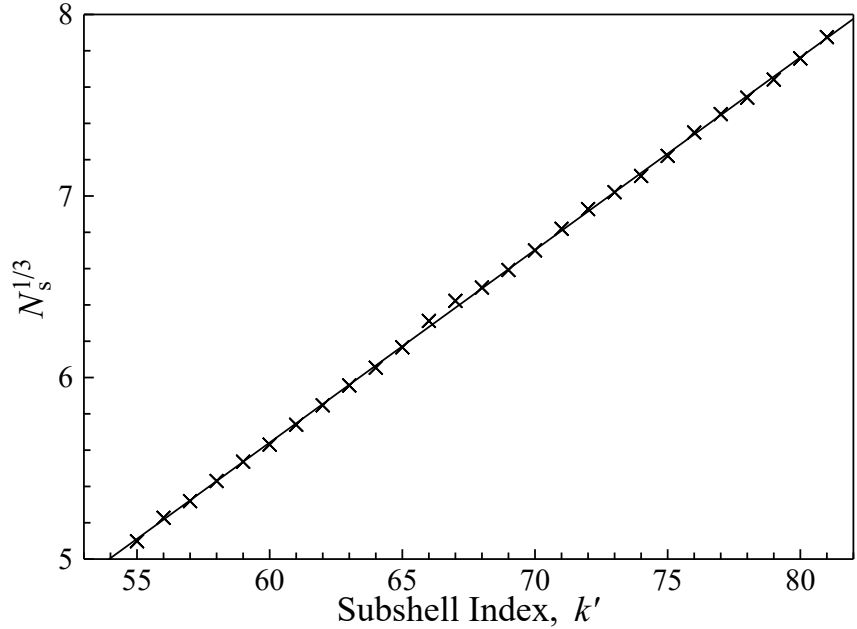

Figure 9: The cube roots of cluster sizes $N_s$ corresponding to subshell closings, determined from the plot in Fig. 6(d) as described in the text.

## 6  Summary and discussion

It is a key result of this work that the systematic inversion procedure described in Sec. 3, developed on the basis of evaporative dynamics, makes it possible to identify universal underlying patterns and extract intrinsic cluster parameters from $(CO_2)_N$ abundance data acquired under a wide range of generation conditions.

In particular, the cluster stability functions and dissociation energies derived from different mass spectra are all in close agreement above $N \approx 130$, independent of the precise source position and nozzle expansion conditions. This confirms the underlying physical assumption that the size-to-size variations in the present mass abundance spectra represent the outcome of clusters undergoing several in-flight evaporation steps.

The analysis presented in this work allowed us to determine accurate size-to-size relative variations of the cluster binding energies, to estimate their absolute magnitude, and to identify the sizes of especially stable clusters. The sequence of these sizes, *i.e.*, the shell and subshell closings, confirms the geometrical nature of $(CO_2)_N$ cluster packing, with a cuboctahedral character for $N \gtrsim 130$. This is consistent with the *fcc* bulk structure (see, *e.g.*, Ref. 15 and references therein), and with electron scattering experiments on neutral $CO_2$ clusters[41].

Implicit in the foregoing discussion has been the assumption that the shell structure and stabilities deduced from the data are characteristic of the neutral rather than cationic $CO_2$ clusters. In other words, the abundance variations primarily derive from cluster evaporation which happens *en route* from the nozzle to the TOFMS rather than post-ionization. This is consistent with the observation[42] that $CO_2$ clusters require more than a few microseconds after (electron-impact) ionization to evaporate, which exceeds their residence time within the TOFMS extraction region. Furthermore, abundance modulations observed in beams of $(CO_2)_N^-$ clusters produced by low-energy electron attachment [20, 22] have periodicities similar to those detected here in the mass spectra of $(CO_2)_N^+$ clusters [15], which suggests that a charge is in any case of minor importance. It should also be mentioned that the contribution to dissociation energies from a Coulomb term is practically negligible outside the low-$N$ range, partly because of the suppression by its size dependence ($\propto N^{-4/3}$) and partly due to the relatively low polarizability of the (non-polar) $CO_2$ molecules.

One observation worth pointing out here is the slow decrease of the amplitude in the variations of the $D$'s with size. This suggests that these variations are caused by evaporations from edges and can be summarized as a nearest neighbor effect. This in turn suggests that although the observed structures in the spectra are generated by evaporative processes, these occur from solid clusters, at least in the final step(s). This is consistent with the bulk phase diagram where no liquid phase is present at low pressures and temperatures.

It is not surprising but perhaps still worth mentioning that the elemental composition of the molecules is not a determining factor for the structure, as can be seen by comparison with the structure of CO observed in Ref. 43.

The structure of $CO_2$ clusters of sizes below 130 remains an interesting open question. Even below the clear onset of cuboctahedral ordering above $N \approx 130$ there is also a notable degree of structure. However, some features appear to evolve gradually with the nozzle expansion parameters (see the ESI), suggesting the presence of structural and phase transformations in this range [44]. In Ref. 20 the cuboctahedral structures were assigned to clusters above $N = 80$. Those clusters were anionic and the difference from our observed lower threshold, maybe due to this difference in the charge state. Interestingly, electron diffraction studies of neutral clusters [41] showed a cubic structure down to $N = 100$, a limit defined by the instrumental resolution, in agreement with the mass spectrometric results for both negatively and positively charged clusters. The precise shell closings were not possible to determine in these studies, unfortunately. The combined experimental and theoretical study in Ref. 44 suggested a somewhat mixed picture with both icosahedral and cubic elements in $N \leq 100$ neutral clusters. A numerical molecular dynamics study[45], also on neutral clusters, indicated the potential co-existence of a metastable icosahedral structure and a stable *fcc* structure over a range of sizes below 100. Adding to this already mixed picture is the observation that for clusters of sizes 50–70, the shell structure seen in the mass abundance spectra was shown to develop on the time scales of the mass spectrometer flight times[42]. Thus identification of shapes and phases of $CO_2$ clusters in this size region requires more study.

## Conflicts of interest

There are no conflicts to declare.

# Acknowledgements

This work was supported by The National Science Foundation of China with grant No. 12047501 and the 111 Project under Grant No. B20063 (R.H.,K.H.), the U.S. National Science Foundation under Grant No. CHE-1664601 (B.S.K., V.V.K, J.W.N.), and by the Swiss National Science Foundation Grant No. 200020_200306 (J.K., R.S.). We thank G.-L. Hou for discussions initiating this work. The figures for this article have been created using the SciDraw scientific figure preparation system 46.

# Shells in $CO_2$ clusters

John W. Niman,[a] Benjamin S. Kamerin,[a] Vitaly V. Kresin,[a] Jan Krohn,[b] Ruth Signorell,[b] Roope Halonen,[c] and Klavs Hansen[c,d]

[a] Department of Physics and Astronomy, University of Southern California, Los Angeles, CA 90089-0484, USA

[b] Laboratory of Physical Chemistry, ETH Zürich, Vladimir-Prelog Weg 2, CH-8093 Zürich, Switzerland

[c] Center for Joint Quantum Studies and Department of Physics, School of Science, Tianjin University, 92 Weijin Road, Tianjin 300072, China

[d] School of Physical Science and Technology, Lanzhou University, Lanzhou 730000, China

Contents

## S-I. Mass spectrum baseline subtraction and peak integration

Three sample time-of-flight spectra, with their time axes converted to corresponding mass, are shown in Fig. S1. The first step in processing them involves identifying and subtracting a constant baseline. As illustrated in Fig. S2, the collected spectra extend far beyond the range shown above. It is safe to assume that the distant points contain no actual cluster signal and therefore represent the baseline. We extract the last thousand points from the record and use their average as the baseline. This value is then subtracted channel-by-channel from the data. A spectrum after baseline subtraction is shown in Fig. S3.

Subsequently, two additional corrections are applied. The first is a Jacobian factor used in a transformation from the time-of-flight variable to the mass variable. This involves a division of the spectrum by $N^{1/2}$. The second one accounts for the cluster size dependence of the photoioniza-tion cross section via an additional division by $N$. The net effect is a pointwise division of the spectra (such as that in Fig. S3) by $N^{3/2}$. The final outcome is illustrated in Fig. 2 in the main text.

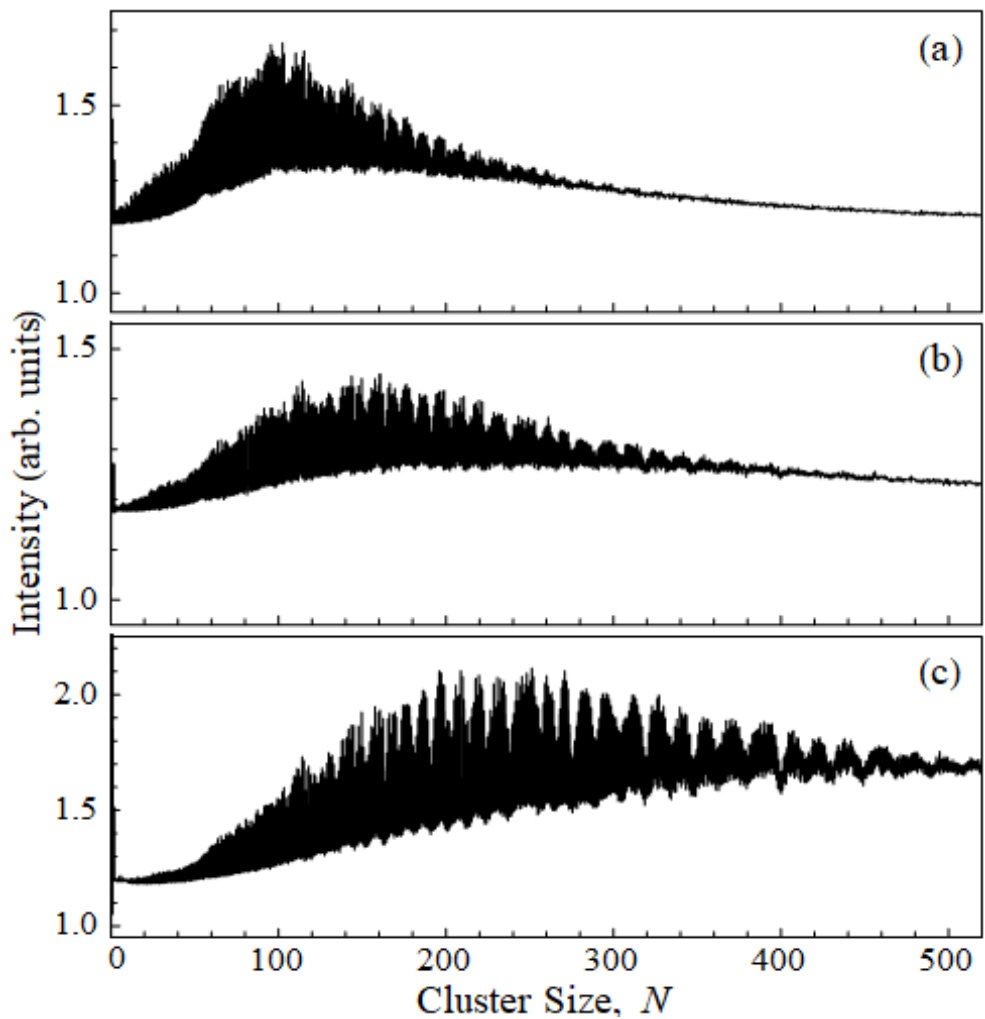

**Fig. S1.** Raw time-of-flight spectra of $CO_2$ clusters for three different experimental conditions. (Spectra are reproduced from data presented in Ref. S1.) The mass spectra in Fig. 2 of the main text are derived from these plots following the procedure described in this section.

Further potential corrections, such as the detector conversion efficiency or other size dependent detection biases, were not included. It is crucial to emphasize that any smooth abundance variations are always removed in the next step, described in Section S-II, and therefore impact neither the stability functions derived there nor any subsequent portion of the analysis. This point follows rigorously from the analysis procedure, and has been verified for the present data.

Following the above corrections, peaks are detected using *Mathematica*'s `FindPeaks` function, with care taken to ensure that exactly one peak is identified for each $(CO_2)_N$ cluster. Fig. S4 shows an example of peaks identified in a baseline-subtracted and corrected spectrum.

Once the $(CO_2)_N$ peak positions are identified, their intensities are determined by numerical integration of a linear interpolation of the data points between the region defined by the midpoints between consecutive peaks. Since there is some variability in where the maxima are determined in the mass spectrum, it is important to normalize the integrated intensity by the distance between the consecutive midpoints. Each integrated intensity is then assigned to an integer value of $N$ corresponding to the cluster size. Variations of this method of peak integration were tried and found to yield similar results. A sample integrated mass spectrum is shown in Fig. S5.

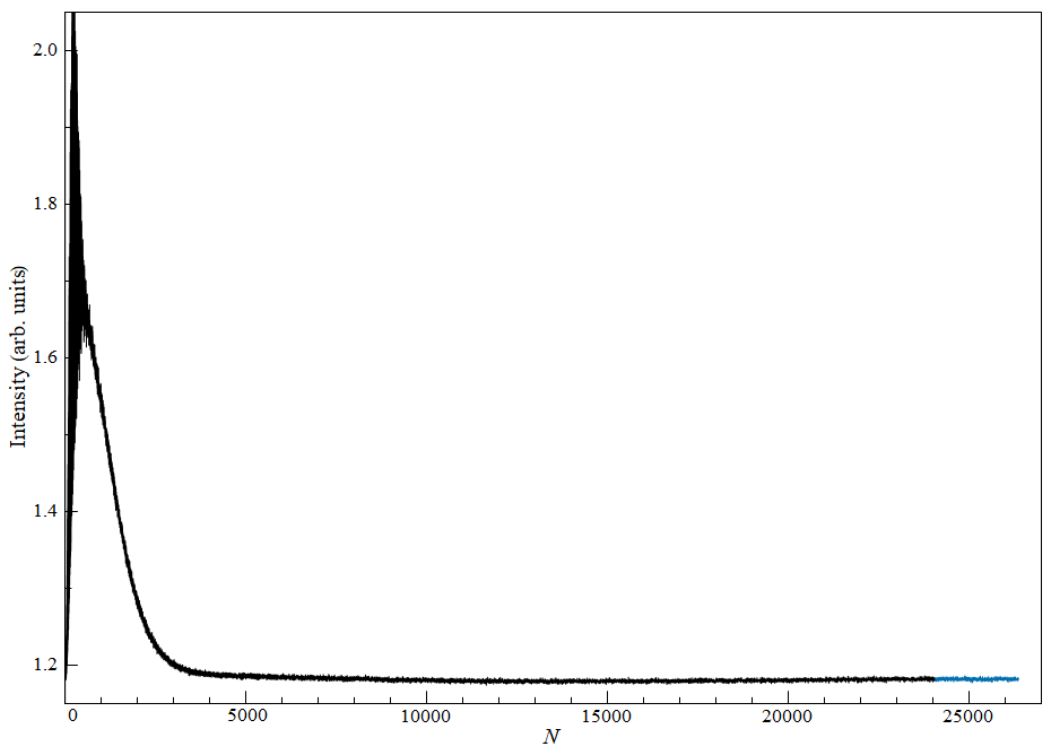

**Fig. S2.** Full time-of-flight mass spectrum of $CO_2$ clusters with points used to construct the baseline function visible in blue at the far end. This plot is an extension of Fig. S1(c). (Spectra are reproduced from data presented in Ref. S1.)

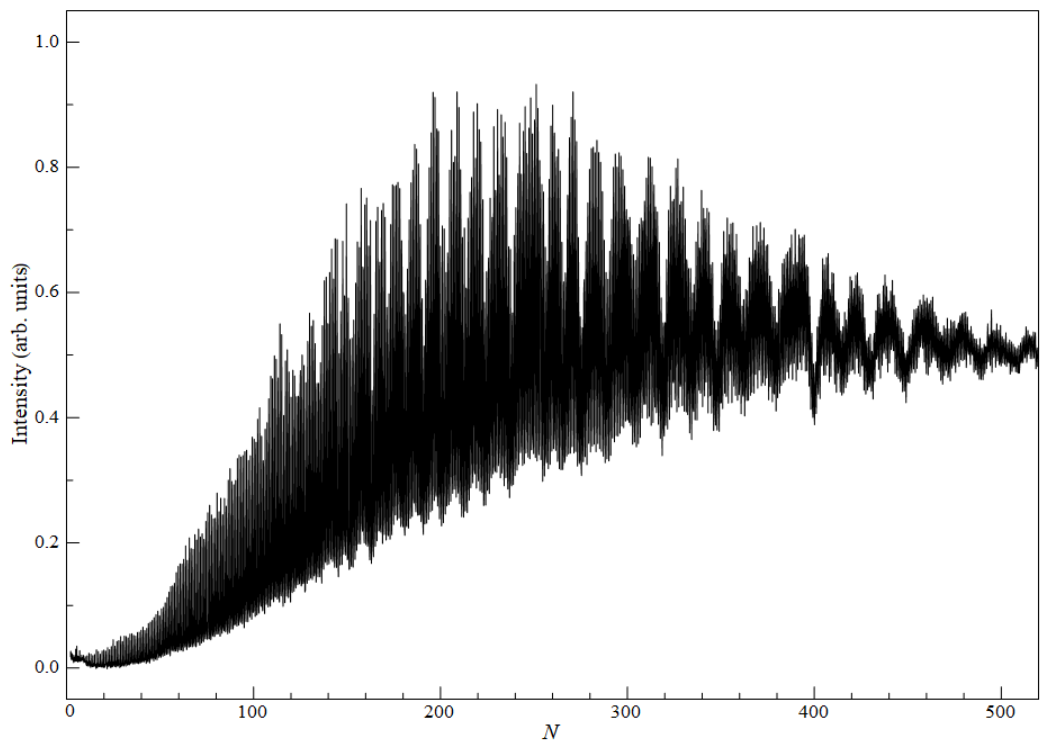

**Fig. S3.** The spectrum from Fig. S1(c) after subtraction of the baseline determined from the region highlighted in Fig. S2. (Spectra are reproduced from data presented in Ref. S1.)

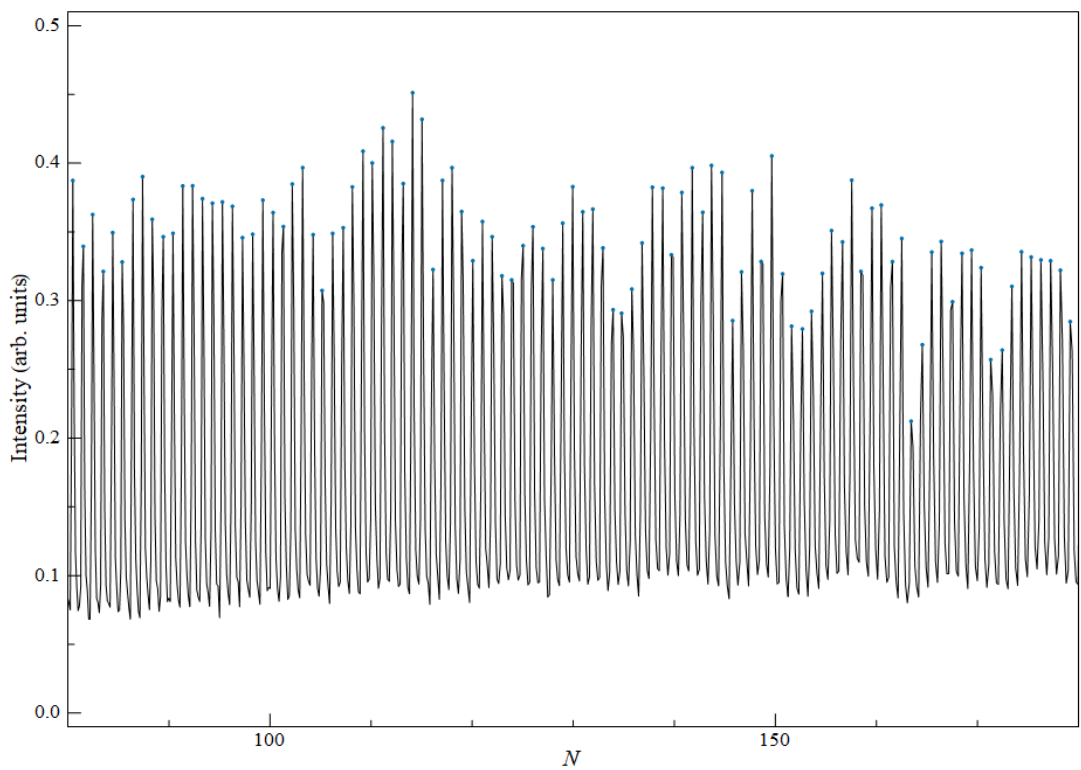

**Fig. S4.** Peaks identified in a segment of the mass spectrum from Fig. 2(c) in the main text.

## S-II. Cluster stability functions

The peak-integrated mass spectra, $I_N$, are used to generate the "cluster stability functions," $(I_N/\tilde{I}_N)$, which exhibit the size-to-size intensity variations deriving from intrinsic cluster properties. The function $\tilde{I}_N$ represents a smooth envelope of the abundance distribution. Since it depends on the source conditions, it is determined separately for each spectrum by means of smoothing the abundance function itself, as described below.

The procedure used here has strong analogies to the determination of shell structure in the field of nuclear physics by means of the so-called Strutinski shell correction method. It is not limited to nuclei and has been applied to studies of cluster shell and supershell structure as well.[S2,S3]

The smoothing is accomplished by convolution of the abundance spectrum with a Gaussian:

$$\tilde{I}_N = \frac{\sum_{N'} I_{N'} \exp\left[-(N'-N)^2 / (2w_N^2)\right]}{\sum_{N'} \exp\left[-(N'-N)^2 / (2w_N^2)\right]}. \tag{S.1}$$

The denominator normalizes the weight to unity. The width is set to $w_N = 4N^{1/3}$. This $N^{1/3}$ variation is selected because it matches that of the structures appearing in the mass spectra. Setting the coefficient to 4 was found to provide sufficiently broad averaging without washing out the size variation of $\tilde{I}_N$. Other smoothing choices are possible, for example the use of spline functions.[S3]

To avoid asymmetric averaging at the high mass end of the spectra, we fit the falling edge of the spectra to an exponentially decreasing function of the form $\alpha e^{-\beta N}$, extend the spectrum and use this for the determination of $\tilde{I}_N$. These extensions are smooth and therefore will not give rise to any spurious signals. We also attempted to fit the tail of the spectra to a pseudo-Voigt function with a sigmoidally varying width parameter,[S4] and found similar results to using the decaying exponential function. The following analysis and the results in the main text utilize the fit to the exponentially decreasing function for simplicity.

A single convolution of this type is not sufficient to remove all traces of the envelope function, and the procedure is therefore repeated twice more, using the preceding $\tilde{I}_N$ as the input spectrum. The iterative process leads to stability functions which oscillate about unity.

Fig. S5 shows such a final envelope function $\tilde{I}_N$ superimposed onto the integrated mass spectrum. This plot is also shown in Fig. 4(c) in the main text, with further examples displayed in other panels of the figure.

Fig. S6 shows an additional series of stability functions calculated for a subset of the experimental data by the pointwise division of $I_N$ by $\tilde{I}_N$.

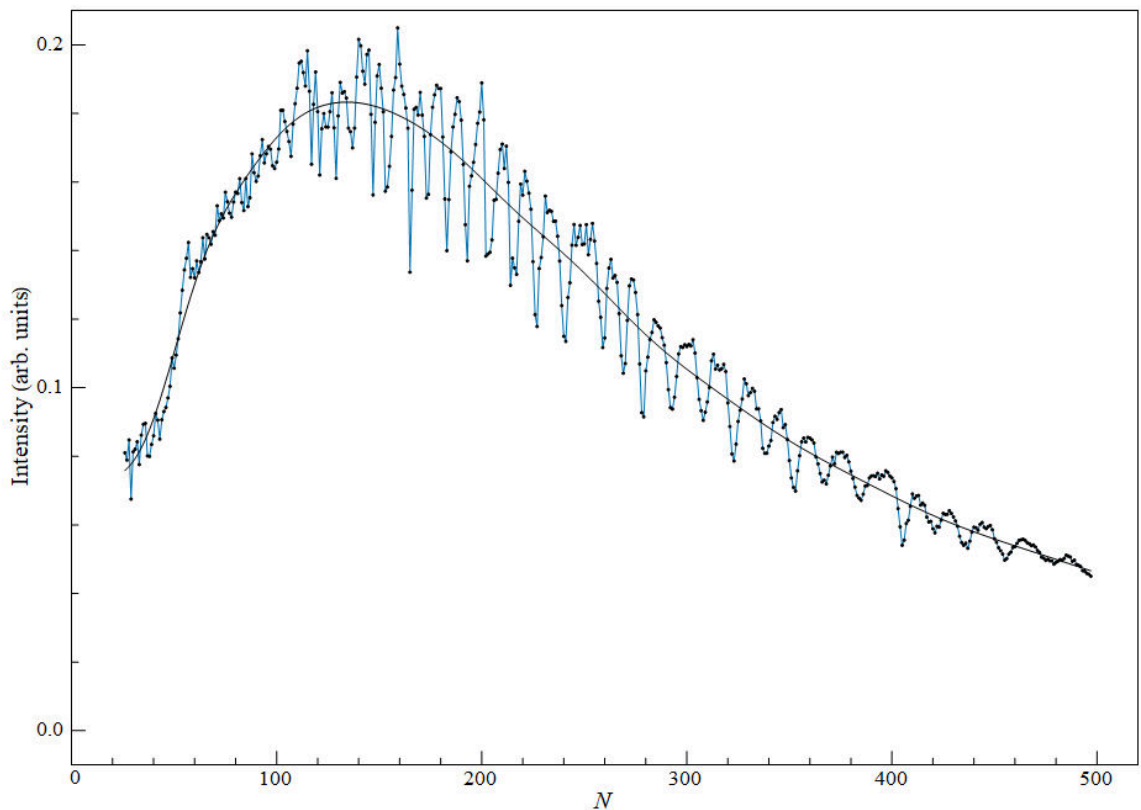

**Fig. S5.** Dots: integrated intensities of peaks identified in the spectrum of Fig. 2(c) of the main text (see also Fig. S4). Smooth solid curve: their envelope obtained by an iterative Gaussian convolution.

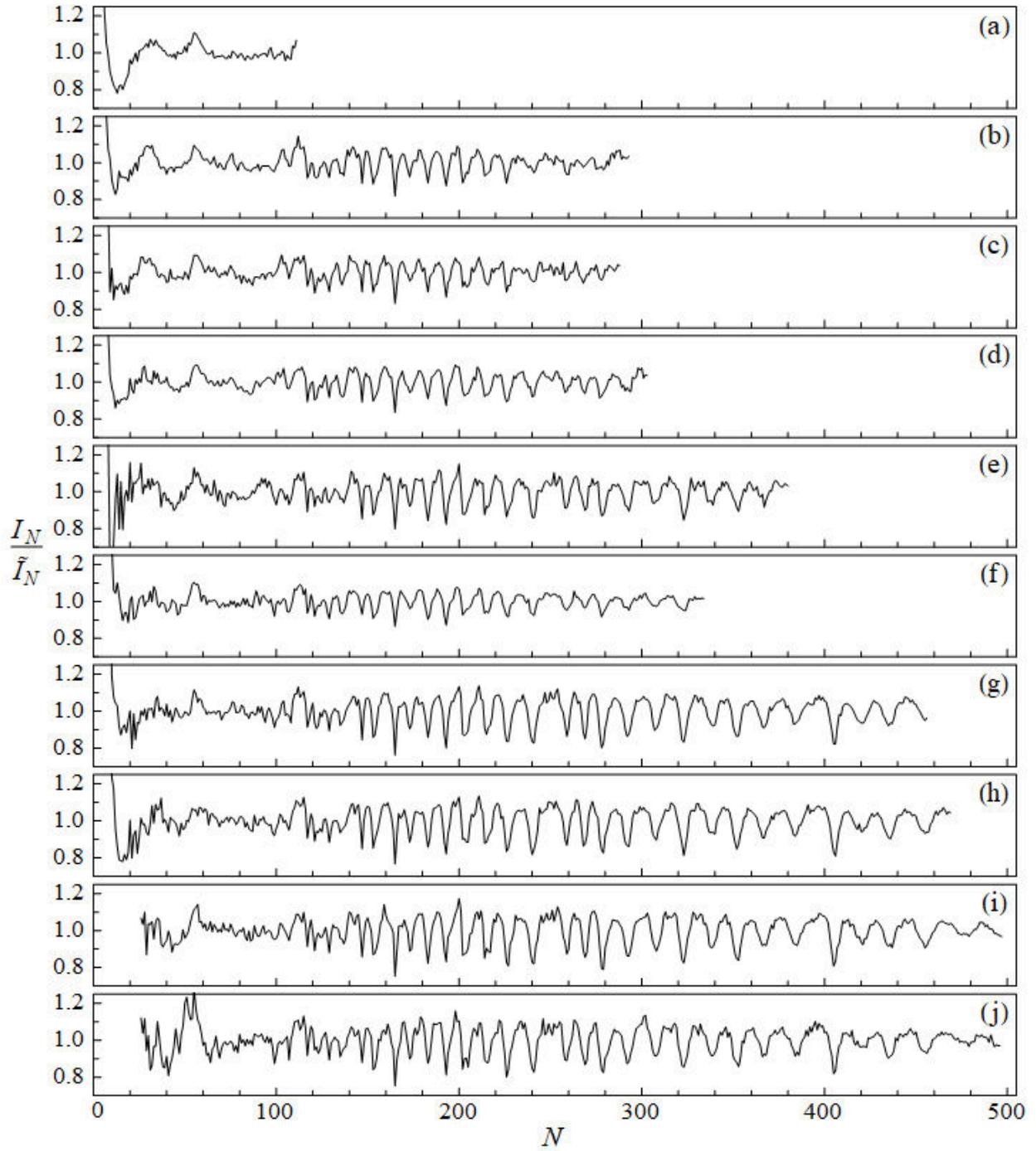

**Fig. S6.** Stability functions for a data set corresponding to 403 mm nozzle-ionization distance and a range of $CO_2$ concentrations in the nozzle expansion: (a) 0.38%, (b) 0.69%, (c) 0.77%, (d) 1.00%, (e) 1.54%, (f) 1.61%, (g) 2.31%, (h) 3.08%, (i) 3.85%, (j) 5.02%.

## S-III. Gspann parameter

The Gspann parameter, defined as $G_N = \ln(\omega_N t)$, derives from a comparison between an isolated cluster's evaporative rate constant and its experimental flight time. In this way, it relates the maximum microcanonical temperature of an evaporative ensemble to its activation energy.[S5,S6]

The rate constant's frequency prefactor can be written as $\omega_N = \sigma_N \Omega$, where $\sigma_N$ is the geometrical cross section for the capture of one $CO_2$ molecule by a cluster of $N-1$ molecules, $\sigma_N = \pi\left[r_0(N-1)^{1/3} + r_0\right]^2$. Here $r_0$ is the effective radius of one molecule [cf. Eq. (4) in the main text], related to the molecular number density $n$ in the bulk by $n^{-1} = (4\pi/3)r_0^3$. The measured density[S7,S8] yields $r_0 \approx 2.2$ Å.

The parameter $\Omega$ can be with good accuracy related to the temperature-dependent bulk vapor pressure $P$ of the cluster material as follows:[S5,S9]

$$P = \Omega\left(\tfrac{1}{8}\pi m k_B T\right)^{1/2} e^{-a/T}. \tag{S.2}$$

Here $k_B$ is the Boltzmann constant, $m$ is the molecular mass, and $\Omega$ and $a$ are fitting parameters. Using the tabulated $CO_2$ pressure data at low temperatures[S8,S10,S11] and plotting $\ln(P/\sqrt{T})$ vs. $1/T$ (Fig. S7) we find from the intercept and the above value of $r_0$ that $\pi r_0^2 \Omega \approx 2.3\times 10^{16}$ s$^{-1}$.

We can now compute the value of the Gspann parameter for each cluster $N$ in each data set. The flight time $t$ is derived from the specific set's distance between the nozzle and the mass spectrometer's ionization region, and the measured cluster beam velocity of 540 m/s. For the size range $10 \leq N \leq 500$ we find that $G_N$ varies between 32 and 35.

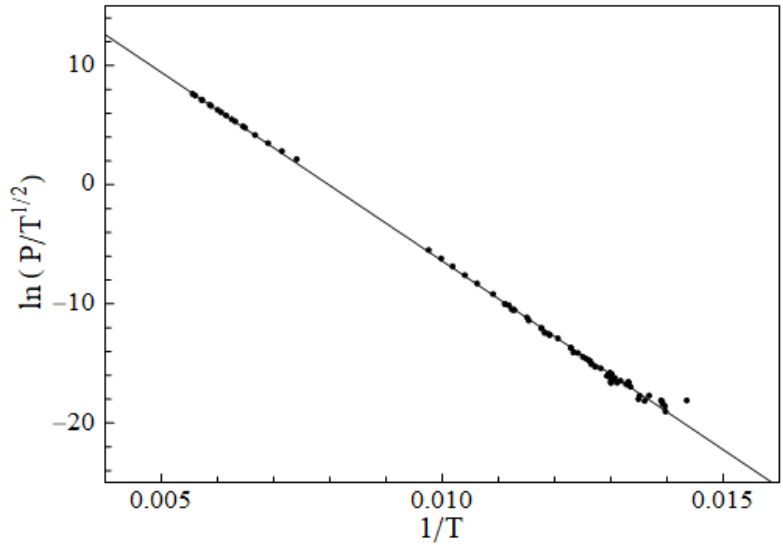

**Fig. S7.** A plot of $CO_2$ vapor pressures in the relevant temperature region.

## S-IV. Heat capacities

Determination of cluster binding energies from the stability functions requires a knowledge of their heat capacities. These are taken from the molar heat capacity[S10] of bulk $CO_2$: $C = 4.6k_B$ per molecule at $T \approx 90$ K.

The temperature estimate above is based on using the relation[S5,S6] $T_N \approx D_N/G_N$ and setting $D_N \approx A$, see Eq. (3) in the main text. While variation of the temperature with cluster size could potentially complicate matters, an earlier analysis for water clusters showed that it can be ignored to a good approximation.[S12]

For use in Eq. (1) in the main text we extrapolate the aforementioned bulk heat capacity to finite clusters sizes by setting it to $C(N–2)$ for a cluster of $N$ molecules. The correction in parentheses corresponds to the subtraction of the six rotational and translational degrees of freedom of the whole cluster.

By taking the average of the precursor ($N+1$ molecules) and detected ($N$ molecules) clusters heat capacities, and remembering that the microcanonical heat capacity of isolated clusters in a beam is one $k_B$ less than the canonical value[S13] we obtain $C_N \approx C(N-\tfrac{3}{2}) - k_B$.

In Eqs. (1) and (2) in the main text this is employed in dimensionless form, i.e., $C_N \approx 4.6(N-\tfrac{3}{2}) - 1$.

## S-V. Dissociation energies

As described in the main text, Eq. (2) is solved recursively by starting with the value of $I_N/\tilde{I}_N$ for the largest cluster in the data set, and proceeding downward in size. The energy ratio $D_{N+1}/\tilde{D}_{N+1}$ is assigned a starting value near unity, and the equation is iterated to find the energy ratios for all the lower sizes. The procedure converges quickly, and within the space of a few tens of molecules the solutions become insensitive to the precise seed value of the energy ratio. For maximum consistency, we select the seed as follows. It is set to values between 0.90 and 1.10 with an increment of 0.0001, and a corresponding set of energy ratios is computed for each of these values. The set which has the smallest average absolute deviation of $D_N/\tilde{D}_N$ from unity for sizes $N \geq 100$ is selected. We found that the seed values optimized in this way did not deviate from unity by more than a couple of percent.

Fig. S8 shows the energy ratios $D_N/\tilde{D}_N$ derived from the data in Fig. S6.

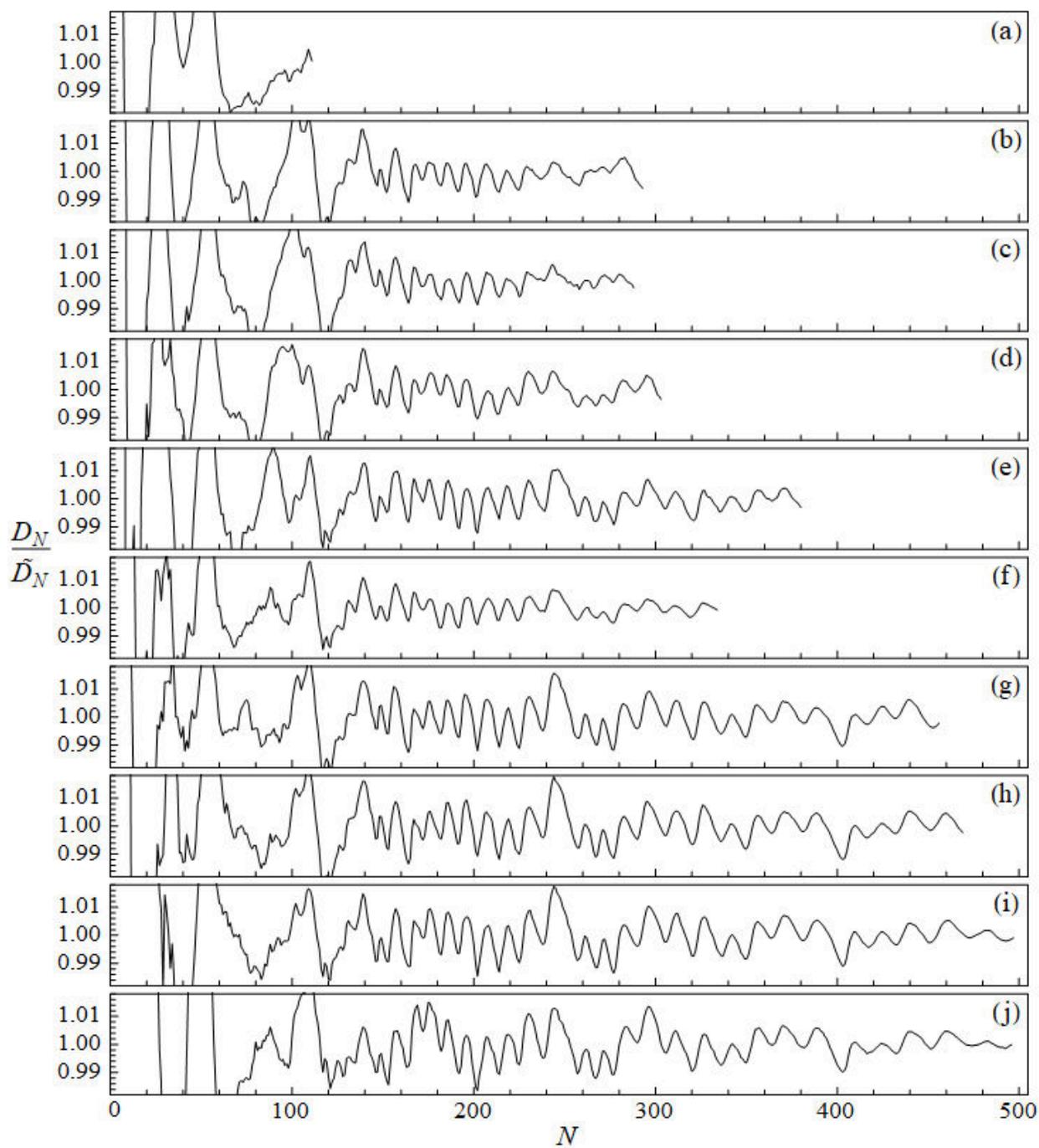

**Fig. S8.** Dissociation energy ratios derived from the stability functions plotted in Fig. S6.

## S-VI. Subshell closings

Fig. S9(a) shows the aggregate set of energy ratios $D_N/\tilde{D}_N$ from Fig. 6(d) in the main text in the region $N \geq 130$. The points of steepest descent are found from this curve by computing the finite difference between successive points, followed by Gaussian smoothing and finally locating the minima, as shown in Fig. S9(b). These cluster sizes, corresponding to sequential facet fillings,[S14] are listed in Table S1 and plotted in Fig. 8 in the main text. These values are the averages of minima candidates obtained by varying the smoothing kernel, and the stated uncertainties derive from the standard deviation of the candidates. A more complicated method involving differentiation of a third order interpolation of the $D_N/\tilde{D}_N$ curve yielded equivalent results.

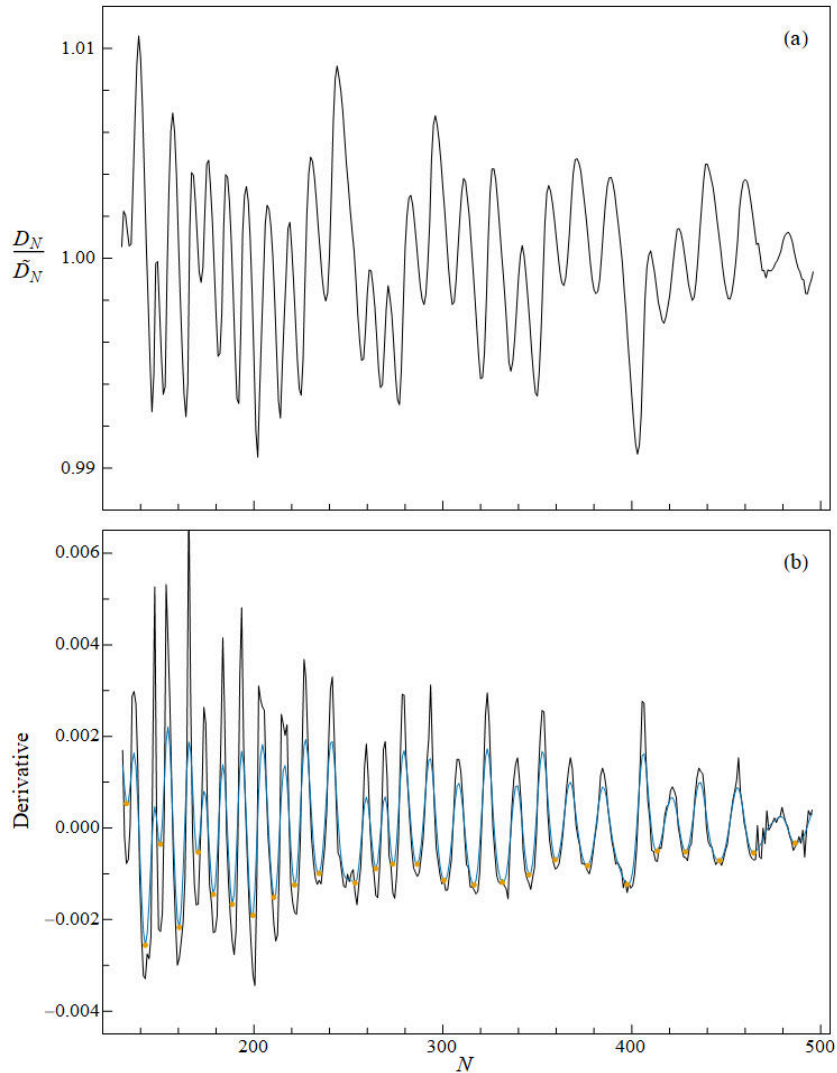

**Fig. S9.** (a) Energy ratios averaged over the complete data set [Fig. 6(b) in the main text]. (b) Derivative of the plot in the top panel (black curve), its Gaussian smoothing (blue curve) and the minimum points of the latter (yellow dots), identifying the points of steepest descent and thereby the subshell closings.

**Table S1.** Cluster sizes with subshell closings as determined from the derivative of the aggregate dissociation energy curve.

| $k'$ | $N_s$ | $k'$ | $N_s$ |
|---|---|---|---|
| 55 | 133 ± 1 | 69 | 287 ± 1 |
| 56 | 143 ± 1 | 70 | 301 ± 1 |
| 57 | 151 ± 1 | 71 | 317 ± 1 |
| 58 | 160 ± 1 | 72 | 333 ± 2 |
| 59 | 170 ± 1 | 73 | 346 ± 1 |
| 60 | 179 ± 1 | 74 | 360 ± 1 |
| 61 | 189 ± 1 | 75 | 377 ± 2 |
| 62 | 200 ± 1 | 76 | 397 ± 2 |
| 63 | 211 ± 1 | 77 | 414 ± 1 |
| 64 | 222 ± 1 | 78 | 429 ± 1 |
| 65 | 235 ± 1 | 79 | 446 ± 2 |
| 66 | 252 ± 3 | 80 | 467 ± 3 |
| 67 | 265 ± 1 | 81 | 488 ± 3 |
| 68 | 274 ± 1 | | |

### S-VII. Geometrical analysis of subshell closings

Refs. S14-S16 discuss the observed oscillating pattern in cluster spectra due to geometrical packing of atoms or molecules. By considering a set of possible polyhedral structures (*fcc* cube, octahedron, decahedron, icosahedron, and cuboctahedron), Negishi *et al*.[S14] concluded that only cuboctahedral clusters represent the mass spectrum of $CO_2$ clusters on a satisfactory level. However, their analysis neglected shapes such as truncated octahedra and Ino (or Marks) decahedra, which are generally plausible structures for larger atomic clusters.[S17-S20]

In what follows we present the geometrical estimates of subshell closings for cuboctahedral, Ino decahedral, and truncated octahedral clusters with $k$ shells. (Sample clusters for each structure are illustrated in Fig. S10.)

**Cuboctahedron.** The total number of monomers in a cuboctahedron with $k$ shells can be written as[S21]

$$N_{\text{cubo}}(k) = \frac{10}{3}k^3 - 5k^2 + \frac{11}{3}k - 1. \tag{S.3}$$

By means of subshell index $k' = Fk$ ($F$ being the number of facets on a shell) the cube root of $N_{\text{cubo}}$ can be developed as a series for large $k'$:

$$N_{\text{cubo}}^{1/3} \approx \frac{(10/3)^{1/3}}{F}(k'-7) + \frac{49}{90^{2/3} Fk}\left(1 + \frac{7}{Fk}\right), \quad (S.4)$$

where the coefficient preceding the first term is about 0.1067 for $F = 14$. The last term of Eq. (S.4) is smaller than 0.05 for the cluster sizes studied here ($k \gtrsim 4$).

**Ino decahedron.** Ino decahedra are also characterized by the shell index $k$ with an additional parameter $p$ (a positive integer). Geometrically, $k$ and $p$ are the numbers of monomers on the edges between (100) and (111) facets and two (100) facets, respectively [see Fig. S10(b)]. The number of monomers in an Ino decahedron is given by

$$N_{\text{Ino}}(k) = \frac{5}{6}k^3 - \frac{5}{2}k^2 + \frac{8}{3}k + p\left(\frac{5}{2}k^2 - \frac{5}{2}k + 1\right) - 1. \quad (S.5)$$

When $p = k$, a regular Ino decahedron has the number of monomers as a cuboctahedron of $k$ shells. But with $F = 15$ instead of 14, the cube root of $N_{\text{Ino}}$ can be expressed to good accuracy by

$$N_{\text{Ino}}^{1/3} \approx \frac{(10/3)^{1/3}}{15}(k' - 7.5). \quad (S.6)$$

Thus the prefactor is about 0.100 and the residual term is about 0.05. However, the energetically more favorable clusters have $p < k$ due to a reduced number of monomers on the costly (100) facets. For such Ino decahedra, the highest order term has again a coefficient of $\approx 0.100$, but the residual terms for the studied cluster sizes are smaller than $\sim 0.025$.

**Truncated octahedron.** The composition of a truncated octahedron differs slightly from the other geometries, and for sake of simplicity an index $n$ ($\geq 0$) is used instead of the shell index $k$.

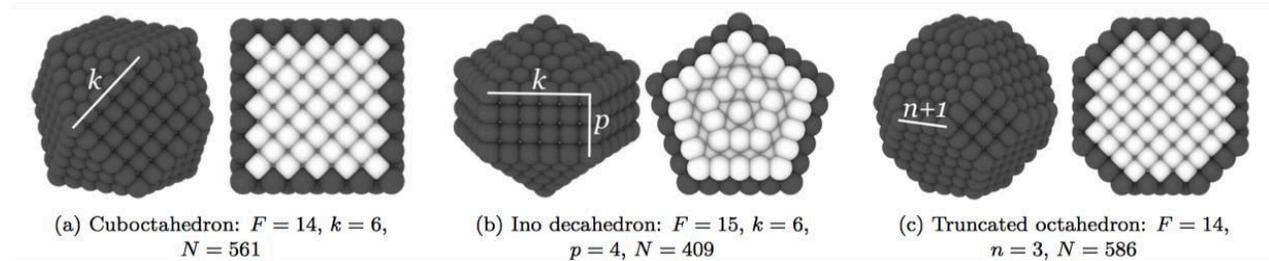

(a) Cuboctahedron: $F = 14$, $k = 6$, $N = 561$

(b) Ino decahedron: $F = 15$, $k = 6$, $p = 4$, $N = 409$

(c) Truncated octahedron: $F = 14$, $n = 3$, $N = 586$

**Fig. S10.** Examples of the three analyzed geometrical structures. The surface monomers are shown as dark spheres and the core monomers (with a coordination number of 12) as bright ones. The complete structure is shown on the left, and the cross-sectional view is given on the right.

The number of monomers in an "$n$th" regular truncated octahedron (TO) is

$$N_{TO}(n) = 16n^3 + 15n^2 + 6n + 1, \tag{S.7}$$

and the number of monomers with a coordination number of 12 (i.e., the number of core monomers) is

$$N_{TO,12}(n) = 16n^3 - 15n^2 + 6n - 1. \tag{S.8}$$

Based on these two equations and a simple geometrical inspection, the complete depletion of monomers from the surface of the $n$th truncated octahedron does not produce the $(n–1)$st octahedron but a cluster with $N_{TO}(n) - 30n^2 - 2$ monomers. Thus extra facets are effectively depleted during a full transition from $n$ to $(n–1)$ closed-shell cluster. After the 14 facets are depleted, the amount of excess surface monomers, $\Delta N$, is

$$\Delta N = N_{TO,12}(n) - N_{TO}(n-1) = 18(n^2 - n) + 5. \tag{S.9}$$

The number of monomers on a subshell of the intermediate cluster can be taken as the average of subshell monomers of the two adjacent closed-shell clusters:

$$\mathcal{N} = \frac{30n^2 + 2 + 30(n-1)^2 + 2}{2 \times 14} \approx 2(n^2 - n). \tag{S.10}$$

Thus the number of effective facets between the intermediate cluster and the $(n–1)$st cluster is approximately $\Delta N/\mathcal{N} \approx 9$. The total number of facets between two regular truncated octahedra is $F = 23$. For truncated octahedral clusters the cube root of $N_{TO}$ is approximately

$$N_{TO}^{1/3} \approx 0.110(n' + 7) + \frac{0.02n + 0.07}{n}, \tag{S.11}$$

where the index $n'$ corresponds to $k'$-14. Thus Eq. (S.11) has a $(k'–7)$ factor similar to Eqs. (S.4) and (S.6). Again the residual term is very small for the relevant cluster sizes ($n \gtrsim 2$).

The geometric analysis for each of the considered structures [Eqs. (S.4), (S.6), and (S.11)] suggests that the subshell index $k'$ is scaled by a factor of $F/2$. This scaling can be explained as resulting from the imperfect indexing of the subshells of the smallest closed-shell cluster. This is demonstrated for the cuboctahedral clusters of $k' \leq 14$ in Fig. S11. Indeed, as shown in the main text, the positions of the subshell closings follow the "cuboctahedral indexing" of $k'–7$.

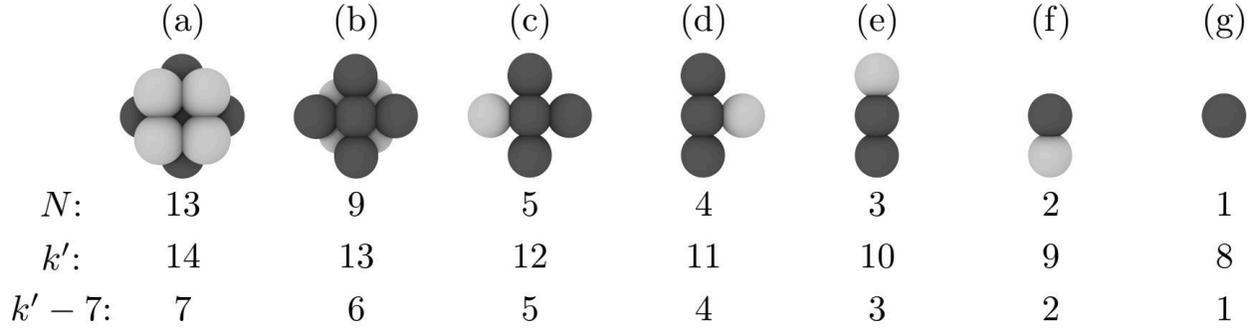

| | (a) | (b) | (c) | (d) | (e) | (f) | (g) |
|---|---|---|---|---|---|---|---|
| $N$: | 13 | 9 | 5 | 4 | 3 | 2 | 1 |
| $k'$: | 14 | 13 | 12 | 11 | 10 | 9 | 8 |
| $k' - 7$: | 7 | 6 | 5 | 4 | 3 | 2 | 1 |

**Fig. S11.** Subshell indexing of small cuboctahedral clusters. (a) Closed-shell cluster with 13 monomers corresponding to $k=1$ (and $k'=14$). Removal of the four monomers on a (100) facet, indicated as light gray, results in a cluster with 9 monomers and subshell index $k' = 13$ [shown in (b)]. This procedure can be repeated for the subsequent clusters and their facets (being either layers or single monomers) until a single monomer remains (g). The shown sequence of subshell configurations (and the shown values of $k'$) demonstrates that the shifting of $k'$ by 7 results in correct subshell indexing, as for $N=1$ the index $k'-7 = 1$.

### S-VIII. Simulation of a peeling-off process

To test the predictive power of the prefactors of the highest-order terms for open-shell structures, we study Lennard-Jones (LJ) clusters with a simple simulation strategy. In our model, the surface monomers on a closed-shell cluster are peeled off one monomer at a time, and after each removal the cluster's energy is minimized using a conjugate gradient algorithm. The monomer to be removed is primarily determined by its coordination number, and secondarily by the minimized energy of the cluster after the monomer removal. Thus the most undercoordinated monomer, whose removal results in the lowest configurational energy, is selected. Note that in this scheme only the surface monomers of the original closed-shell cluster are removed: after a complete depletion of surface monomers a new closed shell remains. The routine is then repeated for the new uncovered closed-shell cluster. A very similar approach was recently used to study the structural motifs of Au clusters.[S20]

It should be also noted that the energy minimization scheme employed here conserves the basic geometry: during the minimization the cluster is not able to collapse into the global minimum (or any other) structure. In the case of LJ clusters consisting of less than 1000 monomers, the global minima exhibit predominantly icosahedral structures.[S22] This is the reason for employing a simple energy minimization instead of an extensive search in phase space using the Hamiltonian.

The stabilities of open-shell clusters are assessed based on the minimized cluster energies $E_N$ with the standard parameters[S21] $\Delta$ and $\Delta_2$:

$$\Delta(N) = \frac{E_N - NE_b}{N^{2/3}} \tag{S.12}$$

and

$$\Delta_2(N) = E_{N+1} + E_{N-1} - 2E_N . \tag{S.13}$$

For a LJ crystal the bulk energy of the *fcc* lattice per monomer, $E_b$, is about -8.6 in standard LJ energy units. The most stable clusters should be located at the local minima and maxima of $\Delta(N)$ and $\Delta_2(N)$, respectively.

The selected starting closed-shell structures are as follows: a cuboctahedron with 561 monomers ($k = 6$), an Ino decahedron with 409 monomers ($k = 6$ and $p = 4$), and a truncated octahedron with 586 monomers ($n = 3$).

The obtained energy parameters $\Delta$ and $\Delta_2$ for cluster sizes between $N = 85$ and 409 are shown in Figs. S12(a) and S12(b), respectively. As the peaks appearing in $\Delta_2$ are more distinct than the minima in $\Delta$, these peak positions are considered as the subshell closings $N_s$. The obtained values of $N_s^{1/3}$ as a function of their order of appearance (corresponding to the subshell index $k'$) are shown in Fig. S11(c). Based on the peeling-off calculations, the geometrical predictions for the highest-order term of $N$ are able to accurately capture the oscillation between the most stable open-shell cluster sizes. However, fluctuations in the values of $N_s$ preclude an accurate analysis of the residuals.

We reiterate that the present model analysis is helpful for identifying the relationship between geometrical structures and the corresponding periodicities of shell and subshell closings. But it should not be used for a direct comparison of cluster configurations and their relative energies between model and experiment because the intermolecular interactions within $CO_2$ clusters are not sufficiently accurately captured by a coarse-grained LJ model and the peeling-off simulations.

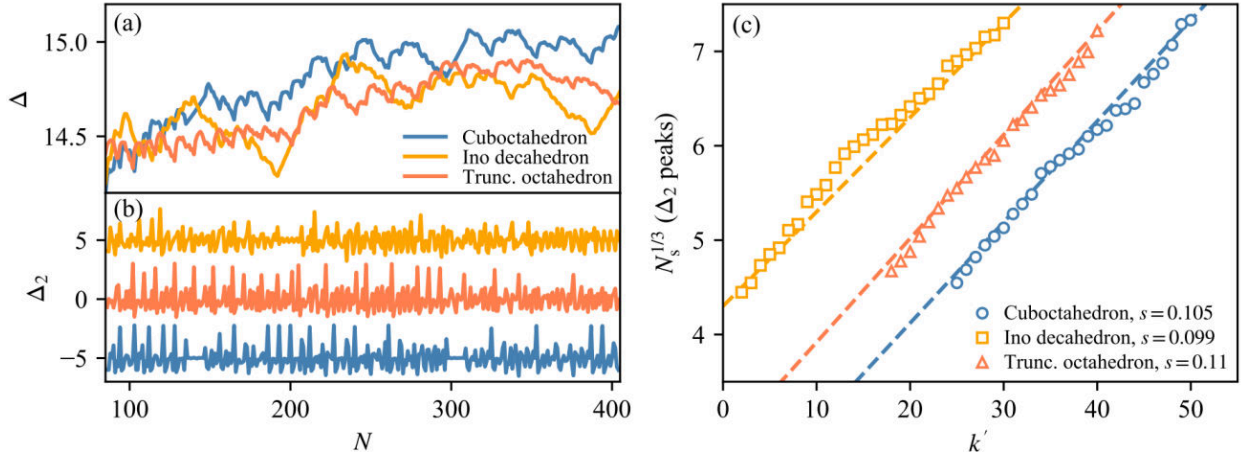

**Fig. S12.** Results obtained from the peeling-off simulations of LJ clusters having either cuboctahedral, Ino decahedral or truncated octahedral geometries. (a) Energy parameter $\Delta$ as a function of cluster size $N$. (b) Energy parameter $\Delta_2$ for the same clusters. For clarity, the lines for cuboctahedral and Ino decahedral clusters are shifted by $-5$ and $+5$ energy units, respectively. (c) Cube root of cluster sizes represented by the peaks appearing in $\Delta_2$, as a function of their order of appearance $k'$. The calculated slopes, $s$, are given in the legend. The points are arbitrarily shifted so that the largest considered $k'$ considered has a value of either 30, 40, or 50. The theoretical predictions ($s_{\text{cubo}} = 0.107$, $s_{\text{Ino}} = 0.100$, and $s_{\text{TO}} = 0.110$) for these geometries are shown as dashed lines (again allocated according to the largest closing).